*Article*

# The Effects of Splayed Lipid Molecules on Lubrication by Lipid Bilayers

Di Jin * and Jacob Klein

Department of Molecular Chemistry and Materials Science, Weizmann Institute of Science, Rehovot 76100, Israel
* Correspondence: di.jin@weizmann.ac.il (D.J.)

**Abstract:** The outstanding lubrication of articular cartilage in the major synovial joints such as hips and knees, essential for the joint well-being, has been attributed to boundary layers of lipids at the outer cartilage surfaces, which have very low friction mediated by the hydration lubrication mechanism at their highly hydrated exposed headgroups. However, the role of spontaneously present lipid splays—lipids with an acyl tail in each of the opposing bilayers—in modulating the frictional force between lipid bilayers has not, to date, been considered. In this study, we perform all-atom molecular dynamics simulations to quantitatively assess the significance of splayed molecules within the framework of lubricating lipid bilayers. We demonstrate that, although transient, splayed molecules significantly increase the inter-membrane friction until their retraction back into the lamellar phase, with this effect more steadily occurring at lower sliding velocities that are comparable to the physiological velocities of sliding articular cartilage.

**Keywords:** hydration lubrication; lipid bilayers; molecular dynamics; biological lubricants; synovial joints; cartilage lubrication

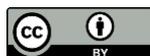

## 1. Introduction

Owing to the advent of an aging society, the quality of life of senior citizens has become growingly important. Osteoarthritis, a disease that is prominently induced by aging, causes pain and reduces mobility in the elderly population. It is projected that in the United States, the number of OA patients will grow to 67 million within the next decade, with more than half of the population of 65 years and above likely to experience the disease. OA is marked by the degradation of the cartilage, and it is a slowly progressive process [1]. A healthy joint has a friction coefficient as low as 0.001 or even lower and is capable of withstanding a pressure up to the order of 100 atm, which protects the cartilage from wear [2]. As the synovial lubrication mechanism fails in the presence of damaged cartilage surfaces, OA patients suffer from significant pain owing to inflammation in joints that frequently endure high friction [3]. It is therefore of paramount importance to understand the mechanism of synovial lubrication. Gaining insights into how biological joints naturally reduce friction will shed light on solutions that can restore joint lubrication and alleviate OA symptoms. Moreover, on a broader scale, an effective biological lubrication therapy will significantly diminish the societal burden of caring for osteoarthritis [4], which currently incurs an estimated annual cost of approximately USD 27 billion in the United States alone [1].

Over the past few decades, a growing body of evidence has demonstrated that hydration lubrication is a fundamental mechanism of cartilage lubrication [5,6]. Hydration lubrication originates from the unique properties of the hydration shell around a charge, i.e., water molecules (so-called hydration water) that are tenaciously packed around a charge through dipole–charge interaction [7]. The duality of the hydration shell, being structurally stable under normal stress and at the same time being sufficiently fluid for dissipating shear stress, is fundamental for its role in biological lubrication. This duality has been demonstrated experimentally from surface force balance (SFB) experiments [7]. In the experiment, shear stress was measured between atomically smooth mica surfaces across sodium solution prepared with ultra-pure water. Under a pressure of greater than 1.5 MPa, the mica surfaces are forced to remain separated by a sub-nanometer distance due to the sodium ions trapped in between them to neutralize the negatively charged mica surfaces. The only water molecules in the separation are the primary hydration shell of the sodium ions, at ca. six water molecules per $Na^+$. Remarkably, the viscosity calculated from the shear stress measurement indicates that the subnanometer-thick hydration shell has an effective viscosity only two orders of magnitude greater than that of bulk water, which is low enough to account for the very low frictional dissipation as such shells compress and slide past each other.

Within the synovial cavity, hydration lubrication is predominantly modulated by lipid molecules that are ubiquitously available [8,9]. Lipids self-organize into the form of liposomes or membranes, coating the cartilage surfaces [10]. The SFB experiments emulate the affine contact at the interfaces between two cartilage surfaces in contact. Each mica substrate is coated by a single lipid bilayer or liposome and shear forces were measured under pressures up to 100 atm over a micron-sized contact area. It has been repeatedly demonstrated that friction coefficients between a variety of phosphatidylcholine (PC) lipid membranes are remarkably low, at $10^{-5}$–$10^{-4}$ under physiological pressures, comparable with that of biological joints [11–15]. PC lipids consist of a hydrophilic zwitterionic headgroup and acyl tails, which leads to bilayers with headgroups exposed to the aqueous environment and the hydrophobic acyl tails interacting to stabilize the bilayer structure. The hydration lubrication scheme is facilitated by the hydrophilic head groups, which contain two oppositely charged groups—the quaternary ammonium group and a phosphate moiety [11–13]. The size of the zwitterionic headgroup allows it to hold, on average, more than 12 water molecules in its hydration shell, effectively providing a sub-nanometer thick water phase between the membranes even under pressure; the repulsive interactions between such shells is termed hydration repulsion [16]. The hydration water phase allows the shear motion to be dissipated mostly in the form of hydrodynamic dissipation, minimizing surface damage [7,16–21].

Although not directly participating in the shearing motion, the role of carbohydrate tails should not be overlooked. The transition temperature TM of the lipid species is typically lower if the carbohydrate tails contain a higher proportion of unsaturated C-H bonds, and it marks the temperature below which the membrane shifts from a liquid phase to a gel phase, when the lipids tails are more aligned and more tightly packed. The packing of the tails controls the overall robustness of the membranes' ability to resist pressure-induced hemifusion, at which point the double-membrane configuration breaks down. The absence of an inter-bilayer water phase leads to the failure of the hydration lubrication mechanism, and consequently potential surface damage under significantly increased friction [11,12,15]. However, the fluidity of the liquid-phase lipid molecules offers a distinct advantage—namely that they can be self-healing during shear [12,22]. In a study of SFB measurements of shear stress between mica surfaces across a liposome dispersion [12], it was discovered that during the shearing motion, unsaturated palmitoyloleioylphosphatidylcholine (POPC) membranes are capable of mending naturally occurring water defects by recruiting lipid molecules from the surrounding environment. From AFM images, POPC membranes become clearly smoother after being smeared by an AFM tip. This effect enables unsaturated lipid membranes to outperform their saturated counterparts, exhibiting lower coefficients of friction, although it is only significant when shearing in the presence of excess liposomes rather than with only pure water in an aqueous environment [12,13]. Clearly, the synovial fluid contains a range of lipid molecules that may each contribute differently to joint lubrication [8,9]. The robustness of the saturated lipids and fluidity of the unsaturated lipids both contribute positively to maintain the hydration lubrication mechanism of lipid membranes at cartilage surfaces. It is thus of considerable interest to understand how the many different lipids in synovial fluid (over 100 of which are known [8,9]) behave, to obtain better insight into their lubricating properties when forming the boundary layer membranes on articular cartilage.

The current study focuses on a common aspect of liquid-phase lipids—their tendency for forming splayed lipids between the membranes—where we consider the lipid boundary layers on articular cartilage, and how such splayed lipids affect the friction between the lipid bilayers during shear in the context of synovial lubrication. Splayed lipids are lipids found at a bilayer–bilayer interface; they straddle across the inter-bilayer water phase with one tail in each of the opposing bilayers. This configuration is metastable as the hydrophilic headgroup rests in the water phase with the hydrophobic tails inserted into the hydrophobic region of the bilayers. It is a transient state more often found in unsaturated lipid membranes due to the flipping of an unsaturated acyl tail protruding over the headgroup region [23,24]. In a stable double lipid bilayer stack, splayed lipids dynamically emerge and disappear under thermal perturbation [24]. Based on previous molecular dynamics calculations, a single POPC lipid molecule transitions into a splay at an average rate of 0.03 $s^{-1}$ when the bilayers are separated at a dehydrated hydration level of 5 water molecules per lipid [25]; double or triple-splayed molecule bundles also occur at lower rates [24]. A single-splayed lipid is metastable for a lifetime of hundreds of nanoseconds if not transitioning to the formation of a lipidic bridge, a fully developed hydrophobic connection between the two membranes [25–28]. By linking the two proximal leaflets, splayed lipids may interrupt the continuity of the inter-bilayer water phase as well as increasing the shear dissipation by promoting direct interaction between the two membranes. The current study aims to investigate the behavior of the splayed lipids when the two bilayers are sheared relative to each other using atomistic nonequilibrium molecular dynamics simulations. This method additionally allows us to quantitatively measure the friction force during shearing in presence of splayed lipid molecules while gaining insights on the dynamic transition of their structures during the process.

## 2. Results and Discussion

Our simulation system is a solid-membrane system designed to emulate the surface force balance experiment. The double bilayer stack contains 128 POPC lipids/monolayer (Figure 1). A bundle of three splayed lipids straddles the inter-membrane water phase and remains stable over the course of 1 μs of equilibration under a pressure of 10 atm. Notably, due to the interaction of the solid slabs with the lipid membranes and the interaction within the splayed lipid bundle, the splayed lipids displayed enhanced stability comparing with the single splayed lipids studied with membrane-only systems in previous works [25,28]. The splayed lipids are expected to fully retract into the lamellar phase in times of the order of microseconds based on existing MD studies of splayed POPC molecule [25,28]. Nonequilibrium simulations were then carried out by setting the solid phases in relative lateral motion (Figure 1d) using a harmonic potential at pulling velocity $v$, and the resulting shear stress $σ(t)$ was recorded during the sliding motion using the GROMACS pull code (see Section 4: Materials and Methods). A range of sliding speeds was tested, varying from 0.1 m/s to 1 m/s, resulting in a final sliding distance of 100 nm to 1000 nm. The lower limit corresponds to ca. 10 times of the size of the simulation box in the lateral direction. Notably, the simulations were performed at sliding velocities and a pressure relevant to the physiological conditions of synovial joints of healthy adults engaged in typical daily activities [6,31].

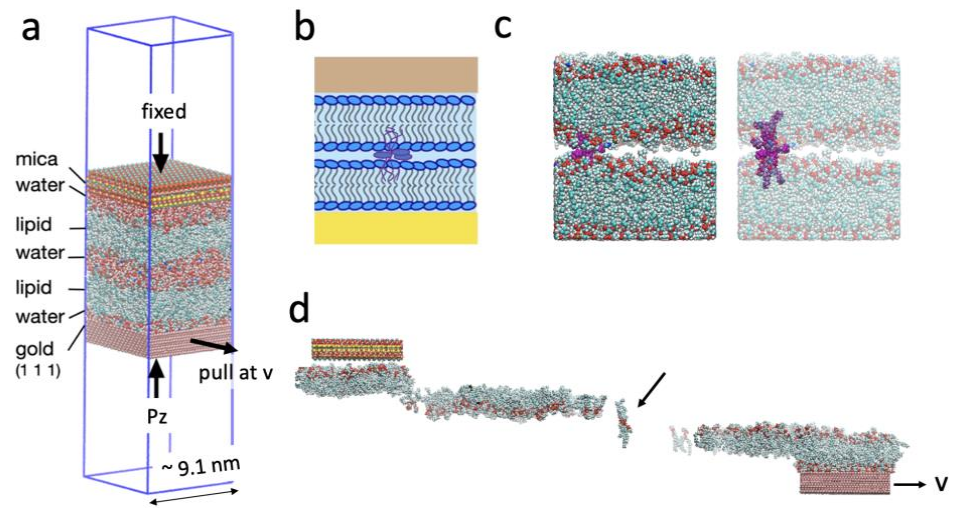

**Figure 1.** (**a**) The simulation configuration replicates the conditions of a surface force balance experiment. (**b**) Schematic of the simulated structure. Lamellar lipids are depicted with headgroups outlined in dark blue and tails in grey. Splayed lipids are outlined in purple. Light blue background: water. Brown: mica surface. Gold: gold surface. (**c**) Snapshots of the simulated structure, omitting the solid phases and water molecules for clarity. The three splayed lipids are labelled in purple. (**d**) A snapshot during shearing, where the splayed lipids are clearly linking the two bilayers (indicated by arrow). The simulations were performed under periodic boundary conditions, equivalent to shearing two infinite planar solid-membrane layers. These layers are constructed of the depicted unit, which repeats infinitely along both the x and y directions. All snapshots within current study are consistently labeled with the same color coding. Blue: nitrogen; gold: phosphorus; red: oxygen; cyan: carbon; white: hydrogen.

The results in Figure 2 demonstrate shear stress traces $\sigma(t)$ that are quickly relaxing to steady-state values, which are positively correlated with the shear velocity for $v \geq 0.5$ m/s, consistent with the governing principle of hydrodynamic dissipation (Figure 2a). An anomaly to this trend is seen at $v = 0.1$ m/s and $v = 0.25$ m/s. At $v = 0.25$ m/s, a distinct and abrupt reduction of frictional stress at ~500 ns is observed. The steady-state shear stress at $v = 0.1$ m/s is higher than that of $v = 0.25$ m/s, contradicting the expected positive correlation between shear stress and shear velocity. To investigate the nature of these anomalous trends, we examined their correlation with the presence of splayed lipids during sliding (Figure 2b). A lipid is determined to be splayed if the two vectors pointing from the two acyl tails to the headgroup are of opposite signs (see Section 4: Materials and Methods), and the fluctuation in $n_{splayed}(t)$ indicates that the splayed lipids can be floppy and revert between being splayed and not splayed. For $v \geq 0.5$ m/s, the number of splayed lipids $n_{splayed}$ drops to zero within 50 ns, and therefore the shear stresses measured at steady state are unrelated to the splayed molecules and arise mostly through hydration lubrication. For $v = 0.25$ m/s, $n_{splayed}$ abruptly dropped to 0 at ~500 ns, simultaneously with the drop in frictional stress. At $v = 0.1$ m/s, $n_{splayed}$ dropped from 3 to 2 at 750 ns, but remained at 2 for the rest of the simulation. As a result, the steady-state shear stress $\sigma_{ss}$ of $v = 0.1$ m/s is greater than that of the $v = 0.25$ m/s case due to the persistent presence of splayed lipids. In all simulations, splayed lipids persevered for some periods of time within the initial 50 ns (Figure 2b). Plotting the maximum shear stress $\sigma_{max}$ found in this time window as a function of the shear velocity, we observe a monotonous trend of positive correlation. These observations strongly suggest that the presence of splayed lipid molecules contributes significantly to the inter-membrane frictional stress measured.

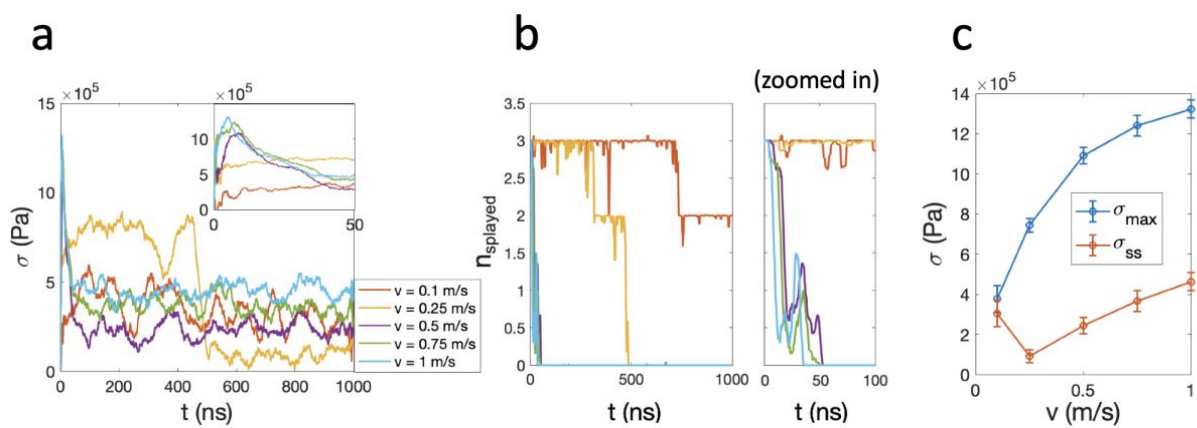

**Figure 2.** (**a**) Frictional stress $\sigma$ measured during the shearing motion at various velocities. (**b**) Left: number of splayed molecules as a function of time, color-coded corresponding to the traces in (**a**). Right: the same figure zoomed in at t < 100 ns. (**c**) The maximum shear stress $\sigma_{max}$ and steady-state shear stress $\sigma_{ss}$ as a function of shear velocity. Error bars indicate the standard deviation of the corresponding $\sigma(t)$ trace at t > 600 ns.

It is speculated that the acyl tails of the splayed molecules, by partially pinning the opposing leaflets, lead to dissipation as different interfaces are sheared, as shown in Figure 1d. To demonstrate this, in Figure 3, the dynamics of the inter-bilayer lipids of the $v$ = 0.25 m/s simulation are visualized by a time series of snapshots, and, on the right, the number of splayed lipids $n_{splayed}(t)$ and shear stress $\sigma(t)$ are plotted with the same x-axis to illustrate their correlation. There are direct visual cues of correlation between the splayed molecules and the friction force. The simulation was initiated with three splayed molecules, which are then pulled by the lower leaflet at the onset of the sliding motion. Over the course of 500 ns, the upper proximal monolayer follows the motion of the lower proximal monolayer through the linkage of the splayed lipids. As the two interphase leaflets (from different bilayers) are coupled, the relative motion was partially dissipated by intra-bilayer tail–tail dissipation, as demonstrated in the snapshot of Figure 1d. At around 500 ns, the splayed lipids retract back into the bilayer under the instability introduced by sliding. Evidently, the shear stress rapidly declines simultaneously and remains at a much lower level attributed to hydration lubrication. This sudden change in lubrication scheme can also be observed from the motion of the upper bilayer, which was initially dragged by the lower bilayer due to the large inter-bilayer friction but comes to a sudden halt when the splayed molecules retract.

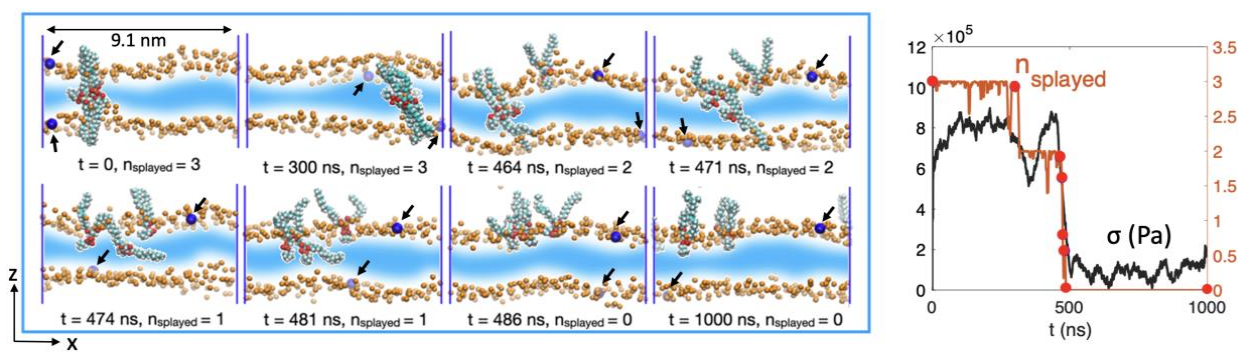

**Figure 3.** (**Left**): Snapshots of the splayed lipids and its correlation with the shear stress measured during a shearing motion at $v$ = 0.25 m/s, where only the midplane monolayers are shown. Depth cueing is applied, with a light color implying greater distance from the viewer. Orange particles: POPC headgroup oxygen (O21 in CHARMM36 forcefield [29,30]). Monolayers' lateral motion is effectively indicated by oxygen atoms labelled in blue and indicated by black arrows—the bottom

bilayer follows the forced rightward motion of the gold slab while the top bilayer slides by the splayed lipid linkage. Notably, the transformation of the splayed lipids back into the lamellar state is accompanied by a significant reduction in shear stress, resulting in the top bilayer becoming static at around $t$ = 474 ns. (**Right**): The corresponding number of splayed lipids and the frictional stress measured as a function of time. The snapshot time points from the right are marked by red markers in the time trace of $n_{\text{splayed}}(t)$.

We have observed from Figure 3 that when the two bilayers move relative to each other, the perpendicular orientation of splayed lipid molecules with respect to the plane of the membrane is disturbed. As the two bilayers move relative to each other at decreasing velocities, the effect of lipid diffusion (with a diffusion coefficient of $D_{\text{lipid}}$) becomes increasingly dominant over the instability introduced by the lateral sliding, and the splayed lipids exhibit a tendency to persist at lower velocities as demonstrated by the case of $v$ = 0.1 m/s. Comparing the time scales of the two processes, namely $\tau_{\text{translation}}$ as the time scale of the splayed molecules being translated by the sliding of the lower bilayer, and $\tau_{\text{diffusion}}$ as the time scale of the splayed lipids diffusing within the leaflet plane, allows us to estimate the critical velocity below which the splayed molecules tend to persist, even during the shearing motion and contribute to the dissipation. From $\tau_{\text{translation}}/\tau_{\text{diffusion}} = (L/v)/(L^2/D_{\text{lipid}}) = D_{\text{lipid}}/Lv > 1$, we obtained $v \lesssim 0.1$ m/s, with $D_{\text{lipid}} \approx 0.1$ nm$^2$/ns, evaluated from the simulations using the GROMACS gmx msd function, and with the characteristic length scale of the splayed lipids $L$ being ca. 1 nm. This limit is consistent with the observation that in our simulation of $v$ = 0.1 m/s, the splayed lipids remained stable over 1 μs. It is of interest that this limit of $v \approx 0.1$ m/s is comparable with physiological sliding velocities of articular cartilage at joint surfaces. This is seen by considering that the maximum angular velocities of lower limb joints is ca. 360°, or $\omega \approx 2\pi$/s (at a gait speed of ca. 1 m/s), obtained by examining healthy adults, corresponding to a cartilage sliding velocity of ca. $\omega r_{\text{femoral}}$ where $r_{\text{femoral}}$ is the radius (≈2 cm) of a femoral head in the knee joint [31].

## 3. Conclusions

This study investigates the role of splayed lipids in lubrication by lipid bilayers through all-atom molecular dynamic simulations. While splayed lipids have been extensively studied in the context of membrane fusion, their role in lubrication has not been explored until now, despite the evident relevance of lipid membranes in synovial lubrication and the prevalence of bilayer stacks at cartilage interfaces, where splayed lipids are known to emerge stochastically. For the first time, our results revealed that the splayed lipids play a significant role in modifying the hydration lubrication mechanism between the lipid membrane boundary layers. In the presence of splayed molecules, the two proximal leaflets at the inter-membrane midplane become coupled through the linkage by the splayed molecules; this reverts the sliding frictional dissipation from the inter-membrane water phase, as in the classic hydration lubrication mechanism, to be partially dissipated through the intra-bilayer tail–tail interaction, significantly increasing the overall dissipation. Previous MD simulations indicated that a single splayed molecule is unstable with a life span of ~300 ns [25], while in our simulations a splayed molecule bundle of multiple splayed lipids appears significantly more stable, with a life span of O (μs) even under shear stress at velocities $v \leq 0.1$ m/s, a physiologically relevant sliding velocity of articular cartilage. It is also relevant to SFB experiments where the sliding velocities are of the order of μm/s. Splayed lipids spontaneously occur within lipid membrane stacks through the thermally invoked flipping motion of lipid tails, usually unsaturated (so in the liquid phase rather than gel phase), and its relevance to membrane lubrication has not been investigated previously. While the current study is a first attempt to solve the problem through a computational framework, it is of relevance to investigate the role of splayed lipids experimentally. As the tendency to form splayed lipids is an intrinsic property determined by the composition of a lipid membrane, this may shed light on the design of lipid additives as biological lubricants [2].

## 4. Materials and Methods

*4.1. Equilibrium Simulations*

The simulation system comprises two 1-palmitoyl-2-oleoylphosphatidylcholine (POPC) bilayers with 128 lipids per monolayer leaflet with a hydration level of 20 water molecules per lipid, as built by the CHARMM-GUI generator [30,32–34]. We used the three-site TIP3P water model [35], and the OH bonds are constrained with SETTLE constraints [36]. For the lipids, we used the CHARMM36 all-atom parameters [29]. Bond lengths were constrained using the LINCS algorithm [37]. Lennard-Jones interactions were cut off at 1.0 nm. The electrostatic interactions were treated with the Particle Mesh Ewald (PME) method [38,39] with a grid spacing of 0.12 nm, a fourth-order spline, and a real-space cut-off distance of 1.1 nm. All simulations were executed with the GROMACS 2018.4 package. The simulation time step is 2 fs. The system was first equilibrated in a NPT ensemble, where the periodic boundary condition is applied in all directions. A velocity-rescaling thermostat was applied with T = 298 K and a time constant of 0.1 ps. The Berendsen barostat was applied with anisotropic pressure coupling, where $P_{x,y}$ = 1 bar and $P_z$ = 10 bars, similar to the experimental condition. The time constant is 2.5 ps and the compressibility is $4.5 \times 10^5$ bar$^{-1}$. The system was equilibrated for 100 ns.

The splayed lipid molecules were induced following the method in Ref [25]. Solid slabs of mica and gold (111), both generated using the CHARMM-GUI generator based on INTERFACE FF, were placed respectively above and below the equilibrated membranes. The thickness of both solid slabs is 2 nm, which is tested to be sufficiently rigid under the applied pressure. Finally, vacuum space is added to the top and bottom of the structure, which totals 25 nm in box height, to reduce interaction between the solid phases across the periodic boundary in the z-direction. The box height is fixed by setting the isothermal compressibility in z to be 0 bar$^{-1}$, which prevents the vacuum space from collapsing. A normal constant compression force of 53 kJ mol$^{-1}$nm$^{-1}$ (≈88 pN) is applied to the center of mass of the mica slab, resulting in a pressure of 10 ± 1 atm, similar to the condition of the SFB experiment.

*4.2. Non-Equilibrium Simulations*

The non-equilibrium simulations were performed with the pull code of GROMACS following the established methods of existing studies [40,41]. To shear the membranes under planar solid confinement, the cross-sectional area is fixed under the NVT ensemble. The mica surface is fixed in all three directions using the freeze group option. The y coordinates of the gold surface are fixed to prevent lateral sliding. In addition, nitrogen atoms of the lipid monolayer in immediate vicinity within 1 nm range from the mica surface are fixed. This is to match the experimental observation that liposomes adhere firmly to the mica surface, even in the presence of strong fluid flow [11]. A harmonic potential with a force constant of k = 1000 kJ/(mol·nm$^2$) was applied to the center of mass of the gold slab, pulling the gold slab at constant velocities of 0.1 m/s, 0.25 m/s, 0.5 m/s, 0.75 m/s, and 1 m/s in the x-direction. The resulting shear force is then evaluated as $F_f$ = $-dU/dx = k(vt-x(t))$. The force traces are written at 0.1 ns intervals using the GROMACS pull code and smoothed using a running average filter with a 50 ns window size. Shear stress is evaluated as $\sigma = F_f/A$, where A is the cross-sectional area of the $9.10 \times 9.16$ nm$^2$ system.

*4.3. Estimating the Number of Splayed Lipids*

Coordinates of the end carbon atoms and the oxygen atom connecting the two acyl tails were exported from the simulations using GROMACS 2018.4 gmx traj function and analyzed in MATLAB 2022b. A splayed lipid is identified if the vectors pointing from the glycerol groups to the two ending carbon atoms have different signs and each with a magnitude longer than 1 nm. The $n_{splayed}$(t) curve is then smoothed over 5 ns intervals.

**Author Contributions:** Conceptualization, D.J. and J.K.; methodology, D.J.; formal analysis, D.J.; data curation, D.J.; writing—original draft preparation, D.J.; writing—review and editing, D.J. and J.K.; visualization, D.J.; supervision, J.K.; funding acquisition, D.J. and J.K. All authors have read and agreed to the published version of the manuscript.

**Funding:** We thank the European Research Council (Advanced Grant CartiLube 743016), the McCutchen Foundation, and the Israel Science Foundation (Grant 1229/20) for financial support; and a Weizmann Institute Cloud Computing grant and the WEXAC and ChemFarm facilities at the Weizmann Institute for computing resources. This work was made possible partly through the historic generosity of the Perlman family.

**Data Availability Statement:** Data will be made available upon request.

**Conflicts of Interest:** The authors declare no conflict of interest.